\documentstyle[12pt]{article}
\begin{document}
\renewcommand{\thepage}{ }
\begin{titlepage}
\title{
\hfill
\vspace{1.5cm}
\\Breakdown of the Fermi Liquid picture in one dimensional
fermion systems:\\
connection with the energy level statistics\\
}
\author{
R. M\'elin, B. Dou\c{c}ot and P. Butaud\\
{}\\
{CRTBT-CNRS, 38042 Grenoble BP 166X c\'edex France}}
\date{}
\maketitle
\begin{abstract}
\normalsize
Using the adiabatic switching of interactions, we
establish a condition
for the existence of electronic quasiparticles in a
Luttinger liquid.
It involves a characteristic interaction strength
proportional to
the inverse square root of the system length.
An investigation of the exact energy level separation
probability
distribution shows that this interaction scale also
corresponds to a
cross-over from the non interacting behaviour to a
rather typical
case for integrable systems, namely an exponential
distribution.
The level spacing statistics of a spin $1/2$, one
branch Luttinger
model are also analyzed, as well as the level
statistics of
a two coupled chain model.
\end{abstract}
\end{titlepage}

\newpage
\renewcommand{\thepage}{\arabic{page}}
\setcounter{page}{1}
\baselineskip=17pt plus 0.2pt minus 0.1pt

The field of strongly correlated electron systems has
recently stimulated
interesting discussions which are sometimes challenging
some more traditional ideas on the many body problem.
For instance, Anderson has proposed that the low energy
properties of a
two dimensional Hubbard model are not properly
described by a Fermi
liquid theory \cite{1}.
In a recent paper \cite{2}, he emphasizes that
this question requires a
non perturbative
treatment, and a careful consideration of boundary
conditions.
As a consequence of the difficulty of the problem,
a lot of effort has been recently dedicated to
numerical investigations
either with Monte Carlo methods or exact diagonalizations \cite{3}.
However, the available sizes remain quite small, and the interpretation
of these results is often delicate.
A rather different approach has been proposed
\cite{4} recently with the hope
to develop new tools for extracting more information from
finite
systems studies. These works have shown that for a
large class of low
dimensionnal strongly correlated systems, the energy levels exhibit
statistical properties rather well described by
random matrix theory.
For instance, a regime of energy level repulsion
is clearly seen
in most investigated cases, at the exeption of
integrable models such
as the nearest neighbor or the $1/r^{2}$ interaction
Heisenberg
spin chain.
Such a behavior has been extensively discussed in the
context of
quantum chaos.
More precisely, it has been verified that many time
reversal
symmetrical classicaly chaotic systems generate a
spectrum in good agreement
with the Gaussian Orthogonal ensemble predictions \cite{5}.
By contrast, simple integrable systems yield in general uncorrelated
energy levels and the usual exponential distribution
for energy
level spacings \cite{6}.
\medskip

In this paper, we are investigating some possible
connections between
simple physical properties of an interacting Fermi system,
such as
the existence of long lived electronic quasiparticles
and the
energy level
distribution. Intuitively, if the energy levels of the
interacting
system keep a simple one to one correspondence with
those of
the non interacting system, we expect on one hand
Fermi Liquid Theory
to be valid, and on the other hand the statistical
properties of
the spectrum to remain qualitatively similar as for
the free
electron case. The interpretation of random matrix
behaviour
is not straightforward. It may simply indicate that
a model is non
integrable. For a normal Fermi Liquid, electronic
quasiparticle
are expected only at low energies compared to the
Fermi energy.
Furthermore, already for the particle-hole phase spaces, interactions
induce new collective modes, such as the zero sound,
and the idea of
a one to one correspondence with the non interacting
gas does not
hold for the whole spectrum. Clearly, it would be very
interesting
to see if the spectrum of a normal Fermi Liquid exhibits
some
features which would distinguish it from a random matrix Hamiltonian.
However, this would likely require an intensive
numerical effort
(since best candidates would be at least two dimensionnal systems).
For the sake of simplicity, and the motivation of doing
analytical calculations, we have concentrated in this work
on a one dimensional model, namely the Luttinger model \cite{7},
which is
integrable at any coupling strength. Interestingly,
this feature
holds for any system length \cite{8}.
Furthermore, it provides a good example of a non-Fermi liquid, which
can be viewed as a non translation invariant fixed point
for many interacting systems in one dimension.
\medskip

This paper is organized as follows.
A first part investigates the condition for the
existence of electronic
quasiparticles, using the adiabatic generation of
eigenstates.
An existence condition is established, from the combined requirement of
having a negligeable generation of non adiabatic components
and absence of decay.
This criterion is satisfied if the interaction strength is
less than a
constant divided by the square root of the system length. As expected, no
quasiparticles are
found for an infinite system at any finite value of the interaction
parameter. This result is also rederived from a simple a
nalysis
of the single particle Green's function for a finite
system.
The second part is devoted to the study of the level
spacing distribution
as the interaction is gradually increased.
We show that the typical interaction scale locating
the departure from
the highly degenerate non interacting system towards a more generic
integrable model with a Poisson distribution is the same as the
previous one.
So, for this simple situation, noticeable change in the energy level
distribution is reflected by the disappearance of electronic quasiparticles.
Then, the last two sections of this paper are dedicated to variants
of this model, namely in the spin $1/2$ case and forward scattering
only, for both one and two coupled one dimensionnal systems. A brief
conclusion summarizes our results.

\section {Adiabatic switching on of interactions}

A formal way to generate quasiparticles in an interacting Fermi liquid
is to apply the Landau switching on of interaction procedure,
namely to start from a free particle added above
the Fermi sea, and
to switch on interactions adiabatically.
The corresponding time dependent hamiltonian is:
\begin{equation}
H = H_0 + V_0 e^{\epsilon t},
\end{equation}
where the interactions term $V_0$ is switched with a rate $\epsilon$.

Provided it is successfull,
this procedure establishes a one to one correspondance between
the
free gas excitations, and the dressed excitations of the Fermi liquid,
namely, the quasiparticles.
For a Fermi liquid, the validity condition of this procedure is
\cite{9}:
\begin{equation}
\label{eq:0}
\Gamma(\epsilon_k) \ll \epsilon \ll \epsilon_k,
\end{equation}
where $\epsilon_k$ is the energy of the quasiparticule, with respect
to the Fermi surface, and $\Gamma(\epsilon_k)$ is the decay rate
of the quasiparticule.
For a normal Fermi liquid, one can show \cite{9} that
$\Gamma(\epsilon_k) \simeq \epsilon_k^{2}$.
At small energies, $\Gamma(\epsilon_k) \ll \epsilon_k$, so
that it is possible to choose a rate $\epsilon$ to perform the
switching on procedure.

The aim of this section is to investigate under which
conditions
the switching on procedure is valid in a one dimensional
Luttinger
liquid.
We shall henceforth exhibit an inequality similar to
equation (\ref{eq:0}) for the rate $\epsilon$
in the case of a Luttinger liquid.

\subsection{Introduction}
We first wish to sum up some results concerning the
formalism
developped in \cite{8}. This will also permit us to fix the notations,
which shall be used in the rest of the paper.

\medskip

The fermions are on a ring of perimeter $L$, with periodic boundary
conditions, so that the wave vectors are quantized
($k=\frac{2 \pi}{L}n$,
with $n$ an integer).

As we treat only
low energy properties of a spinless, one dimensional
Fermi gas, the curvature of the dispersion relation
may be neglected.
The two linear branches in the dispersion relation
emerging from each
extremity of the Fermi surface are extended to
arbitrary energies.
This linearized model is the Luttinger gas model,
which hamiltonian is:
\begin{equation}
H^{0} = v_F \sum_{k p} (pk - k_F) : c_{kp}^{+} c_{kp}
:,
\end{equation}
where $v_F$ is the Fermi velocity and $p=+1$ or $-1$ labels the
branch (right or left).
We shall also use the real space field
$\psi_p^{+}(x)$
associated to the
right (left) free fermions.
Furthermore,
$c_{k p}^{+}$ is the Fourier transform of $\psi_p^{+}(x)$:
\begin{equation}
c_{k p}^{+} =
L^{-1/2} \int_{-L/2}^{L/2}
\psi_p^{+}(x) e^{i k x} dx
{}.
\end{equation}
Notice that the sign of the phase factor is not arbitrary, but
is chosen such as right moving fermions with a positive wave
vector propagate to the right.

Because of the presence of an infinite number of fermions
in the
ground state, the density operators
\begin{equation}
\rho_{q p} = \sum_{k} :c_{k+q,p}^{+} c_{k,p}:
\end{equation}
have anomalous commutation relations (Schwinger terms):

\begin{equation}
[\rho_{q p}, \rho_{-q' p'}] =
- \frac{L p q}{2 \pi} \delta_{p p'}
\delta_{q q'}
{}.
\end{equation}
They may consequently be used to build a set of
boson creators
$a_q^{+}$ ($q \ne 0$).
To handle the real space bosonic field, one needs to define
\begin{equation}
\label{eq:2}
\Phi_p(x) = p \frac{\pi x}{L} N_p - i
\sum_{q \ne 0} \theta(pq) (\frac{2 \pi}{L |q|})^{1/2}
e^{iqx}a_q.
\end{equation}
The $q=0$ modes correspond to charge and current
excitations.
Their algebra
involves the unitary ladder operators $U_p$ constructed in
\cite{8}.
They act only in the $q=0$ sector, and increase
by one the charge on the p branch.
The complete form of the bosonic fields, including
the $q=0$
modes, is:
\begin{equation}
\theta_p(x) = \bar{\theta_p} + \Phi_p(x) + \Phi_p^{+}(x)
,
\end{equation}
where $\bar{\theta_p}$ is the phase conjugate to $N_p$.
\medskip

We shall also use the important relation to pass from a
real space boson description to a real space fermion
description:

\begin{eqnarray}
\label{eq:3}
\Psi_p^{+}(x)
&=& L^{-1/2} e^{-i p k_F x} :e^{-i \theta_p(x)}:\\
\nonumber
&=& L^{-1/2} e^{- i p k_F x}
e^{-i \Phi_p^{+}(x)} U_p e^{-i \Phi_p(x)}
{}.
\end{eqnarray}
Expressed on this new basis, the free hamiltonian
becomes:

\begin{equation}
H^{0} = v_F \sum_{q \ne 0} |q| a_q^{+} a_q +
v_F  \frac{\pi}{L} (N_R^{2} + N_L^{2})
,
\end{equation}
where $N_R$ ($N_L$) denote the number of right (left)
 moving fermions
added above
the vacuum state. In terms of charge $N=N_R+N_L$ and current
$J=N_R-N_L$ variables,
the energy of the charge and current excitations is:
$v_F \frac{\pi}{2 L}(N^{2}+J^{2})$.

\medskip

Note that the action of the boson creation operators and of
the ladder operators on the ground state generates a basis
of the Hilbert space.
The completeness may be shown \cite{8} by comparing
the generating
functions of the degeneracies (ie the finite temperature
partitions functions) for both the free electrons
basis and the
boson basis.
The notation for the kets of the second basis is:
\begin{equation}
|\{N_p\},\{n_q\}\rangle =
\prod_p (U_p)^{N_p} \prod_{q \ne 0}
 \frac{(a_q^{+})^{n_q}}{(n_q !)^{1/2}}
|0\rangle
\end{equation}

\medskip
We now briefly describe the formalism to deal with
interactions.
The two-particle interactions term is written as:
\begin{equation}
H^{1} = \frac{\pi}{L} \sum_{p q} V_q \rho_{q p}
\rho_{-q -p}.
\end{equation}

For simplicity, our treatment does not include
interactions between
fermions lying on the same side of the Fermi surface.
Only $g_2$ interactions are relevant in the physics
we shall develop.
One important feature of the interactions $V_q$ is
that they are
cut off for impulsions greater than the inverse
of a length scale $R$. We shall use the following
expression
of $V_q$ (for $q < 1/R$):
\begin{equation}
\label{eq:30}
V_q = V ( 1- (q R)^{\alpha})
{}.
\end{equation}
The intensity of the interactions is
parametrized by $V$, and the
shape of $V_q$ is parametrized by $\alpha$.
The bosonized form of the interaction Hamiltonian
$H^{1}$ is:
\begin{equation}
H^{1} = \frac{\pi}{2 L}
(v_N - v_F) N^{2}
+ \frac{\pi}{2 L}
(v_J-v_F) J^{2}
+  \sum_{q > 0}
q V_q (a_q^{+} a_{-q}^{+} + a_q a_{-q})
\end{equation}
The total hamiltonian is diagonalized
by the following Bogoliubov transformation:

\begin{equation}
b_q^{+} = \cosh{\varphi_q} a_q^{+} - \sinh{\varphi_q}
a_{-q},
\end{equation}
where the angle $\varphi_q$ is defined as:
\begin{equation}
\tanh{2 \varphi_q} = - \frac{V_q}{v_F}.
\end{equation}
The total hamiltonian reads, after the Bogoliubov
transformation:
\begin{equation}
H = E_0 + \sum_{q \ne 0} \omega_q b_q^{+} b_q
+ \frac{\pi}{2 L}(v_N N^{2} + v_J J^{2}).
\end{equation}
The effect of the interactions is to
give a non zero ground state energy:

\begin{equation}
E_0 = \frac{1}{2} \sum_q (\omega_q - v_F q)
,
\end{equation}
where
\begin{equation}
\label{eq:1}
\omega_q=(v_F^{2} - V_q^{2})^{1/2} |q|
{}.
\end{equation}

Interactions also shift the energies of the oscillators
from $v_F |q|$ to $\omega_q$.
Finally, charge and current excitations acquire different
velocities $v_N = v_S e^{-2 \varphi}$
and $v_J = v_S e^{2 \varphi}$.
In these relations, $\varphi$ is the infrared limit of
$\varphi_q$ and the
sound velocity $v_S$ is related to the infrared limit of
the dispersion relation (\ref{eq:1}):
\begin{equation}
v_S = lim_{q \rightarrow 0}
(v_F^{2} - V_q^{2})^{1/2}
{}.
\end{equation}

In the presence of interactions, one needs to normal
order
the field $\psi_p^{+}(x)$ in terms of $b_q^{+}$ bosons,
which
leads to the following expression of $\Phi_p(x)$ :
\begin{equation}
\Phi_p(x)
=
p \frac{\pi x}{L} N_p
-i \sum_{q\ne 0}
( \theta(pq) \cosh{\varphi_q} - \theta(-pq) sinh{\varphi_q})
e^{iqx} b_q
{}.
\end{equation}

The fermion field reads, in terms of bosons:
\begin{equation}
\label{eq:12}
\psi_p^{+}(x)
=
L^{-1/2}
\exp{\{-\sum_{q > 0} (\frac{2 \pi}{L q})
(\sinh{\varphi_q})^{2}\}}
e^{-i p k_F x}
e^{-i\Phi_p^{+}(x)}
U_p
e^{-i\Phi_p(x)}
{}.
\end{equation}

\medskip

\subsection {Interaction picture for
$c_{kR}^{+}|\{N_p\}\rangle$}
As $t \rightarrow - \infty$, the system is made up of a right moving
fermion,
with an impulsion $k$ added above a Dirac sea
$|\{N_p\}\rangle$,
and interactions are vanishing.
This section deals with the propagation of this state,
$c_{kR}^{+}|\{N_p\}\rangle$,
as interactions
are switched on.

\medskip

The first step is to decompose the state
$c_{kR}^{+}|\{N_p\}\rangle$ into bosonic modes.
The action of $c_{k R}^{+}$ on the vacuum
$|\{N_p\}\rangle$
in the $q=0$ sector
is simply to
increase by one the number of right moving fermions,
by the action of the ladder operator $U_R$.

\medskip

To obtain the action of $c_{k R}^{+}$ in the $q \ne 0$
sectors,
we first Fourier transform $c_{k R}^{+}$ into the
real space field $\psi_{R}^{+}(x)$ for right-moving fermions.

We replace the expression of $\phi_p^{+}(x)$ in (\ref{eq:3})
by its expression  (\ref{eq:2}) in terms of bosonic modes $a_q^{+}$.
The developpement
of the exponential $e^{-i \Phi_R(x)}$ leads then to an
expression of
$c_{k R}^{+}|\{N_p\}\rangle$ as a linear combination of
 bosonic states,
with occupation numbers $\{n_q\}$:

\begin{eqnarray}
\label{eq:4}
c_{kR}^{+}|\{N_p\}\rangle  &=&
\sum_{\{n_q\}} \delta \left( \sum_{q>0}qn_q -
(k-(k_F+\frac{\pi}{L}(2 N_R+1))) \right) \\
& & \prod_{q>0} \frac{1}{\sqrt{n_q!}} \left(
\frac{2\pi}{Lq} \right) ^ {\frac{n_q}{2}}
|\{N_R+1,N_L\},\{n_q\}\rangle
{}.
\end{eqnarray}

The delta function insures that only bosonic states with a
total impulsion
equal to $k-k_F-\frac{\pi}{L}(2N_R+1)$ survive in the decomposition.
As no interaction couples the two branches, creating a right-moving fermion
does not generate left moving bosons.

\medskip

The second step is to propagate
the bosonic wave packet (\ref{eq:4}). Instead of dealing
with the
rather complicated superposition (\ref{eq:4}) of bosonic
states,
we focus
on the propagation of a single term $|\{N_p\},\{n_q\}\rangle$.
We shall use the bosonized form of the two-body interaction hamiltonian,
and look for a solution of the time dependent Schr\"{o}dinger equation:

\begin{equation}
 i \frac{d |\{N_p\},\{n_q\}\rangle_{int}(t)}{dt} =
H_{int}^{1} |\{N_p\},\{n_q\}\rangle_{int}(t).
\end{equation}
The "int" label stands for an interaction picture.
The initial conditions are:
\begin{equation}
\label{eq:01}
lim_{t \rightarrow - \infty}
|\{N_p\},\{n_q\}\rangle_{int}(t) =
|\{N_p\},\{n_q\}\rangle.
\end{equation}
The bosonic states are propagated under the form of a
coherent state:

\begin{equation}
|\{N_p\},\{n_q\}\rangle_{int}(t) = N({z_q(t)}) e^{-i\phi(\{n_q\},t)}
\prod_{q > 0} e^{-i z_q(t) a_q^{+}a_{-q}^{+}}
|\{N_p\},\{n_q\}\rangle.
\end{equation}
The prefactor $N(z)$ normalizes $|\{N_p\},\{n_q\}\rangle_{int}(t)$:

\begin{equation}
N(\{z_q\}) = \prod_{q>0}
\left( 1 - |z_q|^{2}\right)^{\frac{n_q+1}{2}}.
\end{equation}
To determine the time dependent $\phi(\{n_q\},t)$ and
$\{z_q(t)\}$ functions, we first change $z_q(t)$ into $u_q(t)$, with
$z_q(t) = u_q(t) e^{2iv_Fqt}$, and then identify both sides
of the
Schr\"{o}dinger equation. We obtain first order non linear
differential equations for $\{u_q(t)\}$ and $\phi(\{n_q\},t)$:

\begin{eqnarray}
\frac{du_q(t)}{dt} + 2i v_F q u_q(t)
&=& q V_q(t) (1 - u_q^{2}(t))\\
\label{eq:5}
\frac{d\phi(\{n_q\},t)}{dt}
&=&  \sum_{q>0}(n_q+1) q V_q(t) Im(u_q(t))
{}.
\end{eqnarray}
We have discarded in equation ~\ref{eq:5}
a term depending only on $N$ and $J$, which leads
only to a global phase factor.
Translated in terms of $\phi$ and $z_q$ variables, the initial
conditions (\ref{eq:01}) simply mean that $\phi(t)$ and
$z_q(t)$ are vanishing
as $t \rightarrow - \infty$.
These differential equations describe the propagation of
 a single component of the
wave packet (\ref{eq:5}). The propagation
of the summation is obtained as a superposition
of the different components after propagation:

\begin{eqnarray}
(c_{kR}^{+}|\{N_p\}\rangle )_{int}(t) & = &
\sum_{\{n_q\}} \delta \left( \sum_{q>0}qn_q -
(k-(k_F+\frac{\pi}{L}(2 N_R+1))) \right)\\
& & \prod_{q>0} \frac{1}{\sqrt{n_q!}} \left(
\frac{2\pi}{Lq} \right) ^ {\frac{n_q}{2}}
|\{N_R+1,N_L\},\{n_q\}\rangle_{int}(t).
\end{eqnarray}

\subsection {Adiabaticity condition}
We are looking for a solution of equation (\ref{eq:5}) which
depends only on the variable $s=\epsilon t$, in the small
$\epsilon$ limit. It is possible since the external time dependance
in equation (\ref{eq:5}) involves only $\epsilon t$.
We assume then $u_q(s) = u_q^{0}(s) + \epsilon u_q^{1}(s)
+ O(\epsilon ^{2})$.
Neglecting the $O(\epsilon ^{2})$ terms in equation  $(\ref{eq:5})$
leads to:

\begin{eqnarray}
2 i v_F q u_q^{0}(s) &=& q V_q(s) (1 - u_q^{(0)}(s)^{2}) \\
\label{ad1}
\frac{d u_q^{0}}{ds}(s) + 2 i v_F q u_q^{(1)}(s) &=&
-2q V_q(s) u_q^{(0)}(s) u_q^{(1)}(s)
,
\end{eqnarray}
where:
\begin{equation}
\label{ad3}
V_q(s) = V_q^{0} e^{s}
{}.
\end{equation}
The purely adiabatic solution $u_q^{(0)}(s)$ is given by:
\begin{equation}
u_q^{0}(s) = \frac{i}{V_q(s)}
(- v_F + \sqrt{v_F^{2} - V_q(s)^{2}})
= i \tanh{\varphi_q^{0}(s)}
{}.
\end{equation}
Using this solution in equation (\ref{ad1}) gives the first
 finite
$\epsilon$ correction:
\begin{equation}
u_q^{(1)}(s) = i \frac{v_F}{2 q (v_F^{2} - V_q(s)^{2})}
u_q^{(0)}(s)
{}.
\end{equation}

The adiabatic preparation of eigenstates is achieved if
$|u_q^{(1)}(s)| \epsilon \ll |u_q^{(0)}(s)|$ for $s=0$,
which leads to:
\begin{equation}
\frac{v_F \epsilon}{2 q (v_F^{2} - V_q^{2})} \ll 1
{}.
\end{equation}
This condition depends explicitely on $q$, and is satisfied
for
any value of $q$ if
\begin{equation}
\label{ad2}
\epsilon \ll 4 \pi v_F / L.
\end{equation}

Here, we assume a weak coupling, namely $|V_q| \ll v_F$.
It should be noticed that this upper bound on $\epsilon$ is
a much more restrictive condition than the corresponding upper
bound in equation (\ref{eq:0}) for a Fermi liquid. We interpret this
as a consequence of the fact that the quasiparticules of the
Landau theory are not exact eigenstates of the interacting
system. They are obtained in a situation where the thermodynamic
limit is taken first, whereas the generation of exact eigenstates
would require $\epsilon$ to go to zero as the typical spacing between
energy levels. Our criterion (\ref{ad2}) corresponds to this second
situation. This choice has been motivated by the possibility to
construct the exact eigenstates of a Luttinger liquid.

\subsection{Adiabatic propagation in a Bogoliubov subspace}
The aim of this section is to propagate a fermion during the
switching on procedure. We suppose that the condition
(\ref{ad2}) is satisfied, and we now look for a minoration
of $\epsilon$.
We first search an approximation for the evolution operator
in the limit $\epsilon \ll \frac{2 \pi}{L} v_F$.
At the order $\epsilon^{0}$, the evolution
operator $U_{\epsilon}(0,-\infty)$
realizes the Bogoliubov transformations of angles
$\{ \varphi_q^{0} \}$, corresponding to the rotation of the
basis of eigenstates as interactions were switched on from
zero at time $t=-\infty$ to $\{ \varphi_q^{0} \}$ at time $t=0$.
We shall note $U^{0}(0,-\infty)$ the corresponding part
of the evolution operator. $U^{0}$ must have the property that:
\begin{equation}
U^{0} a_q^{+} (U^{0})^{-1}
=
\cosh{ \varphi_q^{0}} a_q^{+} - \sinh{ \varphi_q^{0}} a_{-q}.
\end{equation}
This equality is verified if $U^{0}$ has the following form:
\begin{equation}
U^{0} = \exp{\{ \sum_{q > 0} \varphi_q^{0}
( a_q^{+} a_{-q}^{+} - a_q a_{-q})\}}
{}.
\end{equation}
To see it, we differentiate each operator $U^{0} a_q^{+} (U^{0})^{-1}$
and $U^{0} a_{-q} (U^{0})^{-1}$ with respect to $\varphi_q^{0}$ and
solve the differential system.

\medskip

However, at higher orders in $\epsilon$,
the evolution operator must take into account the
phase factor $\phi(\{n_q\},t)$, the evolution of which is
given by
the equation (\ref{eq:5}). Assuming
that the
propagation is adiabatic, we approximate $Im u_q(t)$ in (\ref{eq:5})
by $Im u_q^{0}(t)$:
\begin{equation}
Im u_q(t) \simeq  \tanh{ \varphi_q^{0}(s=\epsilon t)}
{}.
\end{equation}

We use the expression (\ref{ad3}) for $V_q(t)$,
and integrate the differential
equation (\ref{eq:5}) for the phase factor $\phi(\{n_q\},t)$.
A constant (infinite)
phase factor associated to the propagation of the ground state
is factored out. Thus, we
obtain the form of the evolution operator in the
adiabatic limit (at order $\epsilon^{0}$ for the operator
$U^{0}$, and at order $1/\epsilon$ for the phases):

\begin{eqnarray}
U_{\epsilon}(0,-\infty) & = &
U^{0}
\exp{i\{\sum_{q>0} \frac{q n_q v_F}{\epsilon}
(\varphi_q^{0})^{2}\}} \\
& = &
\label{eq:7}
\exp{\{\sum_{q>0} \varphi_q^{0}
( a_q^{+}a_{-q}^{+} - a_q a_{-q} )\}}
\exp{\{i \sum_{q>0} \frac{q n_q v_F}{\epsilon}
(\varphi_q^{0})^{2}\}}
{}.
\end{eqnarray}
In the integrations, we have assumed that the interactions are weak,
and the phase factors are given, at the lowest order in $\varphi_q^{0}$.

The rest of this section is devoted to the calculation
and the interpretation of the overlap:
\begin{equation}
\label{eq:13}
F(x-x') \sim
F(x,x') = \langle \{N_p\}|
\Psi_R(x') U_{\epsilon}^{-1}(0,-\infty)
\Psi_R^{+}(x) U_{\epsilon}(0,-\infty)
| \{N_p\} \rangle,
\end{equation}
between the dressed fermions
$\Psi_R^{+}(x) U_{\epsilon}(0,-\infty)
| \{N_p\} \rangle$, and the bare ones:
$\Psi_R^{+}(x') U_{\epsilon}^{-1}(0,-\infty)| \{N_p\}
\rangle$.
To perform it, we use the expression
(\ref{eq:12}) of the field for right moving
fermions, and the approximation (\ref{eq:7}) for the
 evolution operator.
The computation is straightforward, and $F(x,x')$ is the
product of three terms:

1) a phase term
\begin{equation}
N = \exp{\{-i(k_F + \frac{\pi}{L} (2 N_R+1))(x-x')\}}
,
\end{equation}
corresponding to the propagation in the $q=0$ sector.

2) A term corresponding to the left moving bosons normal
ordering in (\ref{eq:13}):
\begin{equation}
G_1 = \exp{\{- \sum_{q>0} \frac{2 \pi}{L q}
(\sinh{\varphi_q^{0}})^{2}\}}
\end{equation}

3) A term coming from the right moving bosons normal
ordering:
\begin{equation}
G_2(x,x') = \exp{\{\sum_{q>0} \frac{2 \pi}{L q}
e^{-iq(x-x'-\frac{v_F (\varphi_q^{0})^{2}}{\epsilon})}\}}
\end{equation}
The result for the overlap is:
\begin{equation}
F(x,x') = \frac{1}{L} N G_1 G_2(x,x').
\end{equation}
The $G_1$ term contains the usual physics of the
 orthogonality
catastrophy \cite{10}.
If $\varphi_q^{0}$ is assumed to be constant between
$q=\frac{2 \pi}{L}$ and $q=1/R$, and zero afterwards,
and if $L \gg R$,
$G_1$ can be calculated as:
\begin{equation}
\label{g1}
G_1 = (\frac{L}{R})^{-\sinh^{2}{\varphi^{0}}}
\end{equation}
In the weak coupling limit, one can deduce the
characteristic interaction scale associated to the
orthogonality catastrophy:
\begin{equation}
\label{adia1}
V_{o.c.} = v_F (\ln{\frac{L}{2 \pi R}})^{-1/2}
{}.
\end{equation}
To obtain the energy scale associated to the $G_2$ term,
we use the relation (\ref{eq:30}) and approximate
the phase as:
\begin{equation}
- \frac{q v_F (\varphi_q^{0})^{2}}{\epsilon}
= - \frac{q (V^{0})^{2}}{4 v_F \epsilon}
+ \frac{ q (V^{0})^{2}}{2 v_F \epsilon} (q R)^{\alpha}
{}.
\end{equation}
The first term is linear in $q$ up to the impulsion
 scale $1/R$.
The second term is associated to smaller impulsion scales.
The formers are relevant for a quasiparticle.
If $k$ is the impulsion of the quasiparticle
with respect to the Fermi level, the energy scale $V_{deph}$
associated to the dephasing is given by:
\begin{equation}
\label{eq:31}
\frac{ k V_{deph}^{2}}{2 v_F \epsilon} (k R)^{\alpha}
= 2 \pi
,
\end{equation}
that is:
\begin{equation}
V_{deph}(k) = (\frac{4 \pi v_F \epsilon}
{k (k R)^{\alpha}})^{1/2}
{}.
\end{equation}
The switching on procedure
shall henceforth be successfull provided the intensity
of interactions $V$ is much smaller than $V_{deph}(k)$,
that is:
\begin{equation}
\label{eq:41}
\frac{k V^{2} (kR)^{\alpha}}{4 \pi v_F}
\ll
\epsilon
{}.
\end{equation}

\subsection{Conclusions}
For the switching on procedure to create a quasiparticle,
the conditions (\ref{ad2}) and (\ref{eq:41}) have to
be simultaneously satisfied, that is:
\begin{equation}
\label{eq:20}
\frac{k V^{2} (k R)^{\alpha}}{4 \pi v_F}
\ll
\epsilon
\ll
\frac{4 \pi}{L} v_F
{}.
\end{equation}
This inequality is satisfied if the following consistency condition is
fullfilled:
\begin{equation}
\label{eq:21}
V \ll \frac{4 \pi v_F}{(k L (kR)^{\alpha})^{1/2}}
{}.
\end{equation}
As we shall see, this condition has a simple interpretation
on the spectrum of the Luttinger model.
At this stage, we should again emphasize that the upper
bound on
$\epsilon$ is more restrictive than in Landau theory.
If we use the
more usual condition that the spread in energy is smaller
than the
average density of the wave packet, equation (\ref{eq:20})
is replaced
by:
\begin{equation}
\frac{k V^{2} (k R)^{\alpha}}{4 \pi v_F}
\ll
\epsilon
\ll
k v_F
,
\end{equation}
and the consistency condition is:
\begin{equation}
V \ll (\frac{4 \pi}{(k R)^{\alpha}})^{1/2} v_F
{}.
\end{equation}
The absence of Landau quasiparticule in the thermodynamic
limit is
then atributed to orthogonality catastrophy, as indicated
by equation
(\ref{adia1}).

\subsection{Comparison with the Green's function}
In this section, we calculate the
Green's function for the finite size Luttinger model:

\begin{eqnarray}
\label{eq:22}
G_R(x,t;x',t') &=&
-i
\{
 \langle \{ N_p \} |
e^{i H (t'-t)}
\psi_R(x')
e^{-iH(t'-t)}
\psi_R^{+}(x)
| \{ N_p\} \rangle
\theta(t'-t)\\
\nonumber
& &
- (x \leftrightarrow x';
t \leftrightarrow t')
\}
\end{eqnarray}

and reestablish the consistency condition (\ref{eq:21}).
Note that in equation (\ref{eq:22}), $|\{N_p\}>$ denotes an eigenstate
of the interacting system.

\medskip

To calculate the Green's function, we use the
expression (\ref{eq:3}) of
the field $\psi_R^{+}(x)$ in terms of the Bose field, and
 normal
order the expression (\ref{eq:22}) of the Green's
function with respect to the bosonic modes $b_q^{+}$.
The computation is straightforward, and the result is:
\begin{eqnarray}
G_R(x,t;x',t')
&=& \frac{-i}{L}
e^{i (k_F+\pi/L) (x'-x)}
e^{i \frac{\pi}{L}(v_N (2N+1) + v_J (2J+1))(t'-t)}
\\
\nonumber
& &
\exp{(-2 \sum_{q>0} \frac{2 \pi}{L q}
(\sinh{\varphi_q})^{2})}
\\
\nonumber
& &
\{ [
\exp{(\sum_{q>0} (\frac{2 \pi}{L q})
(\cosh{\varphi_q})^{2} e^{iq(x'-x)}
e^{-i \omega_q(t'-t)} ) }
\\
\nonumber
& &
\exp{(\sum_{q>0} (\frac{2 \pi}{L q})
(\sinh{\varphi_q})^{2} e^{-iq(x'-x)}
e^{-i\omega_q(t'-t)})}
]
\\
\nonumber
& &
-[x \leftrightarrow x';
t \leftrightarrow t']\theta(t-t')
\}
\end{eqnarray}

The dispersion in the frequencies leads to decoherence after
a time $t_k$.
($k$ is the impulsion of the quasiparticle, with respect to the
Fermi level).
$t_k$ may be estimated in the same way as we did for $U_{deph}$,
and one finds:
\begin{equation}
\label{eq:42}
t_k = \frac{2 \pi v_F}{V^{2} k (kR)^{\alpha}}.
\end{equation}

For a system of size $L$,
the wave packet is stable, provided it can cross
the ring without decoherence:

\begin{equation}
v_F t_k > L
,
\end{equation}
that is:
\begin{equation}
V < (\frac{2 \pi}{k L (kR)^{\alpha}})^{1/2} v_F
{}.
\end{equation}
Up to some numerical dimensionless constants, this criterium
is the same as the consistency condition (\ref{eq:21}) for the
switching on of interactions.

\section{Level statistics of the interacting Luttinger model}
\subsection{Introduction}

We first need to find out a proper sector of the
 Hilbert space,
in which we shall compute the level statistics.
We note $H_{JN}$ the subspace with given current $J$
and charge $N$.

In the free case, the boson basis of $H_{JN}$ can be
 organized as follows:
consider all the sets of occupation numbers $\{n_q^{0}\}$
such as, for all $q$, $n_q^{0}=0$ or $n_{-q}^{0}=0$.
The corresponding states $|\{n_q^{0}\}\rangle$ are
 annihilated
by any pair destruction operator:
$a_q a_{-q} |\{n_q^{0}\}\rangle =0$.
Starting from $|\{n_q^{0}\}\rangle$, and creating pairs
generates a subspace $H_{pairs}(\{n_q^{0}\})$.
A basis of $H_{pairs}(\{n_q^{0}\})$ is made up of all
the states
$|\{n_q^{0}+p_{|q|}\}\rangle$ with arbitrary occupation
numbers
for the pairs $\{p_q\}_{q>0}$.
$H_{NJ}$ is the direct sum of all the
$H_{pairs}(\{n_q^{0}\})$
subspaces.

\medskip

The subspaces  $H_{pairs}(\{n_q^{0}\})$
remain stable under
the action of
the interaction hamiltonian
$H^{1}$, so that they are appropriate to the study
of the levels evolution.

\medskip
We choose $N=J=0$ and
drop the energy term associated to $\{n_q^{0}\}$,
since we always
handle differences between consecutives levels.
The energy levels are given by:

\begin{equation}
E(\{n_q\}) =
\sum_{q>0} 2 v_F q n_q(1 - (\frac{V_q}{v_F})^{2})^{1/2}
\end{equation}
where we use the expression (\ref{eq:30}) for $V_q$.

\subsection{Description of the algorithms}
In this section and the next paragraph, we use
reduce units
for the energies and impulsions: $\omega$ is an
energy divided
by $\frac{2 \pi}{L} v_F$ and q is an impulsion divided
by $\frac{2 \pi}{L}$.

The degeneracies of the Luttinger model are given by:
\begin{equation}
g(\omega) = \sum_{\{n_q\}} \delta (\omega-\sum_{q>0} q n_q)
{}.
\end{equation}
Replacing the $\delta$ function by its integral
representation leads to:
\begin{equation}
g(\omega) = \int_{-L/2}^{L/2} \frac{dx}{L} \prod_{q>0}
\frac{1}{1-e^{iqx}}e^{-i \omega x}.
\end{equation}
Let $g^{(k)}(\omega)$ be the number of different sets of occupation
numbers, having the property that:

\begin{equation}
\omega = \sum_{l=k}^{\omega}l n_l.
\end{equation}
Of course, $g^{(1)}(\omega) = g(\omega)$.
The integral representation for $g^{(k)}(\omega)$ reads:
\begin{equation}
g^{(k)}(\omega)
= \int_{-L/2}^{L/2} \frac{dx}{L} \prod_{q \ge k}
\frac{1}{1-e^{iqx}}e^{-i \omega x}.
\end{equation}
Using the integral representations for
$g^{(k)}(\omega)$, we obtain the following recurences:
\begin{equation}
\label{eq:40}
g^{(k)}(\omega) = \sum_{\nu=k}^{\omega} g^{(\nu)}
(\omega-\nu)
,
\end{equation}
which allows us to numerically compute $g(\omega)$.

\medskip

With a similar recursion,
we may generate all the states of the free Luttinger
model:
the states with an
energy $\omega$ are obtained by adding a boson with an
impulsion $\nu$ on the states with an energy
$\omega - \nu$.

\medskip

As far as the interacting Luttinger model is concerned,
we need to generate all the energy levels
with an energy inferior
as a given cut-off $\omega_0$.
Since there are an infinite number of
levels in the sector under consideration, we need
to introduce such a cut-off to compute the statistics.
We shall then compute the statistical properties of this
set of levels.
If a sufficient number of levels with an
energy inferior as $\omega_0$ has been generated,
the statistical properties are independant on $\omega_0$.
To generate the levels,
we remark that
the frequencies of the oscillators increase with their
impulsion.
So that we successively fill up the
individual oscillator levels, starting
with the smallest frequencies.

\subsection {Level statistics}
\subsubsection{Degeneracies of the free Luttinger model}

Using the recursion relation (\ref{eq:40}), we computed
the
degeneracies of the first 800 levels
of the free Luttinger model.
The asymptotic form of the density of states may be
derived
in terms of initial fermions. The partial degeneracies
for
n-particules n-holes excitations in a one branch model
are:
\begin {equation}
g^{(n)}(\omega)
=
\sum_{\{k_i\}_{i=1...n}}
\sum_{\{k'_i\}_{i=1...n}}
\delta(\omega - \sum_{i=1}^{n} \omega(k_i) -
\sum_{i=1}^{n}
\omega(k'_i))
{}.
\end{equation}
The sets $\{k_i\}$ ($\{k'_i\}$) are the impulsions of the
holes (particules), and are constrained by the Pauli
principle
$k_i \ne k_j$ ($k'_i \ne k'_j$) for all indices $i \ne j$.
This sum is approximated by assuming a constant density
of states, neglecting the Pauli exclusion principle,
replacing
the discrete sum by an integral:
\begin {equation}
g^{(n)}(\omega)
=
\frac{1}{(n!)^{2}}
\int_{0}^{\omega} d\omega_1
\int_{0}^{\omega-\omega_1} d\omega_2
...
\int_{0}^{\omega-(\omega_1+...+\omega_{2n-1})}
d\omega_{2n}
\delta(\omega-(\omega_1+...+\omega_{2n}))
\end{equation}
The multiple integral is readily evaluated and
leads to:
\begin {equation}
\label{XXX}
g(\omega) = \sum_{n=1}^{+\infty} g^{(n)}(\omega)
= \sum_{n=1}^{+\infty}
\frac{\omega^{2n-1}}{(n!)^{2} (2n-1)!}
{}.
\end{equation}
For sufficiently large energies,
the sum may be approximated by its saddle point value,
approximately reached for the following value of $n$:
\begin {equation}
n^{*} = \sqrt{\frac{\omega}{2}}
{}.
\end{equation}
The degeneracy evaluated at $n=n^{*}$ is:
\begin {equation}
g^{n^{*}}(\omega) \sim
\frac{2^{3/4}}{(2 \pi)^{3/2}}
\frac{1}{\omega^{5/4}} \exp{\sqrt{8 \omega}}
{}.
\end{equation}
We computed the summation (\ref{XXX}) in order to test
the accuracy of the saddle point approximation,
which is plotted on figure \ref{degeneracies}.
The exact degeneracies of the Luttinger model reveal to
be inferior as the saddle point asymptotic form,
which is
imputed to the exclusion principle
(figure \ref{degeneracies}).

\subsubsection{Qualitative structure of the spectrum}
The evolution of some energy levels as a function of
the interactions
is plotted on figure \ref{energy levels}.
In this spectrum, we distinguish
two regions:

1) No level crossings are present at sufficiently small
energies
and interactions. The free Luttinger model ($V=0$)
belongs to
this part of the spectrum. In this region,
the statistics are ill defined for they strongly depend on
the energy cut off.

2) If $E$ and $V$ are large enough, level crossings occur,
and level statistics are Poisson
statistics. The convergence of the statistics as a function
of the energy cut off $e_0$ is shown on figure
\ref{convergence}.
Here, we emphasize that these level crossings occur
because the
Luttinger model remains integrable at any value of
the coupling
constant.

To characterize the separation between these two regions of
the spectrum,
the location of the crossings is estimated in the
following way:
as the intensity of interactions $V$ is equal to zero, the spectrum
is made up of equidistant degenerate levels, separated
by an amount
of energy $\Delta E = \frac{2 \pi}{L} v_F$.
As $V$ is turned on, the degeneracies are lifted.
We focus on a single
fan of levels.
All the levels are degenerate if $V=0$, and their energy is
$E^{0} = 2 v_F k$, where $k$ is the total impulsion
of the states.
For a given value of $V$, all the levels lie between
$E_{min}$ and
$E_{max}$. $E_{min}$ is obtained as all the quanta
are in the
smallest energy state (namely $q=\frac{2 \pi}{L} v_F$),
so that:
\begin {equation}
E_{min} = 2 v_F k
(1 - \frac{V_{q=\frac{2 \pi}{L}v_F}^{2}}{v_F^{2}})^{1/2}
{}.
\end{equation}
$E_{max}$ corresponds to a state with one quantum in the highest
$q=k$ state:
\begin {equation}
E_{max} = 2 v_F k (1 - \frac{V_{q=k}^{2}}{v_F^{2}})^{1/2}
{}.
\end{equation}
As the interaction parameter $V$
increases, the levels evolve and
the first crossings occur as the width of the fan
$E_{max} - E_{min}$ is of order $\Delta E$. This condition
 defines
the interaction energy beyond which crossings exist:
\begin {equation}
\label{eq:46}
V^{*} = (\frac{\pi}{k L (k R)^{\alpha}})^{1/2} v_F
{}.
\end{equation}

Via bosonization, the free Luttinger liquid is described
as a set of
harmonic oscillators with commensurable frequencies. As interactions
 are switched on, the oscillator frequencies vary and
become incommensurable.
In \cite{6}, Berry and Tabor show that a system with a finite number of
generic harmonic oscillators does not exhibit level clustering.
It appears that increasing the number of oscillators
with incommensurable frequencies generates clustering.

\subsubsection{Quasi particle destruction and level spacing statistics}
The condition (\ref{eq:46}) separates two regions of the
spectrum.
The same energy scale controls the existence or the absence of
a quasiparticule in a Luttinger liquid, in the sense of
adiabatic
continuation of exact eigenstates. We have thus shown that
the structure of the spectrum of the
finite size Luttinger
liquid is related to the
succes or the failure of adiabatic generation of eigenstates
from the non interacting fermion system.

\subsubsection{Limit $R = 0$}

Consider the case of the two branch
Luttinger liquid with $\frac{1}{R} = + \infty$,
and a constant interaction,
namely, for all $q$, $V_q=V$.
In that case,
all the bosonic modes keep their coherence whatever
the value of $\epsilon$. The decoherence
time $t_k$, given in (\ref{eq:42}), is infinite.
The condition (\ref{eq:41}) associated to the dephasings
is always
verified whatever the value of $\epsilon$.
The only remaining restriction for the switching on
procedure to be
successful is thus:

\begin{equation}
\epsilon \ll \frac{2 v_F}{L}
{}.
\end{equation}

The level statistics are singular in this limit.
The degeneracies of
the fan of levels are never lifted, whatever the intensity
of interactions $V$.
The degenerate levels depend on $V$ in the following
way:
\begin{equation}
E(\{n_q\}) = \sum_{q>0} 2 v_F q n_q
(1 - (\frac{V}{v_F})^{2})^{1/2}
{}.
\end{equation}

However, we note that the overlapp between the eigenstate
 thus
constructed and the state obtained from the action
of the bare
electron operator on the interacting ground state
is vanishing
according to equation (\ref{g1}) since $R=0$.
\section{Level spacing statistics for a spin $1/2$,
one branch Luttinger model.}
The rest of the article is devoted to the study
of some models
derived from the two branch, spinless Luttinger
liquid model.
We begin with the one branch Luttinger model,
with spin $1/2$,
and a $g_4$ interaction. The kinetic energy term
is:
\begin{equation}
H^{0} = \sum_{k \sigma} v_F (k - k_F)
: c_{k \sigma}^{+} c_{k \sigma}:
,
\end{equation}
where the label $\sigma$ denotes the spin
component along the
$z$ axis.
The interaction is given by:
\begin{equation}
H^{4} = \frac{g_4}{2 L}
\sum_{q \sigma}
\rho_{q \sigma}
\rho_{q -\sigma}^{+}
{}.
\end{equation}
The usual spin and charge combinations:
\begin{eqnarray}
\label{eq:45}
C_q^{+} &=&(\frac{\pi}{Lq})^{1/2}
( \rho_{q \uparrow} + \rho_{q \downarrow} )\\
S_q^{+} &=&(\frac{\pi}{Lq})^{1/2}
( \rho_{q \uparrow} - \rho_{q \downarrow} )
,
\end{eqnarray}
have bosonic commutation relations, and the total
hamiltonian
$H = H^{0} + H^{4}$ is diagonal in terms of spin
and charge
variables:

\begin{equation}
H = v_C \sum_{q >0} q C_q^{+} C_q
+ v_S \sum_{q>0} q S_q^{+} S_q
+ v_F \frac{\pi}{2 L} (N_{\uparrow}^{2}
+ N_{\downarrow}^{2})
{}.
\end{equation}
The charge and spin velocities are:
$v_C = v_F + \frac{g_4}{2 \pi}$
and $v_S = v_F - \frac{g_4}{2 \pi}$.

\medskip

The $g_4$ interaction is switched on adiabatically:
\begin{equation}
g_4(t) = g_4^{0} e^{\epsilon t}
{}.
\end{equation}
The evolution operator is:
\begin{equation}
\label{eq:111}
U_{\epsilon}(0,-\infty)
= \exp{\{
-i \sum_{q>0}
\frac{g_4^{0}}{2 \pi \epsilon}
q (n_{C q}-n_{S q})\}}
,
\end{equation}
where $n_{C q} = C_q^{+} C_q$ and $n_{S q} =
S_q^{+} S_q$.
The overlap
\begin{equation}
F(x,x') =
\langle\{N_p\}|
\Psi_{\uparrow}(x') U_{\epsilon}^{-1}(0,-\infty)
\Psi_{\uparrow}^{+}(x) U_{\epsilon}(0,-\infty)
| \{N_p\} \rangle
\end{equation}
is found to be equal to:
\begin{equation}
F(x,x') = \frac{1}{L} e^{i(\frac{\pi}{L}
(2 N_{\uparrow} +1) + k_F)(x-x')}
\frac{1}{( 1 - e^{i \frac{2 \pi}{L}(x-x'
-\frac{g_4^{0}}{2 \pi \epsilon})})^{1/2}}
\frac{1}{( 1 - e^{i \frac{2 \pi}{L}(x-x'
+\frac{g_4^{0}}{2 \pi \epsilon})})^{1/2}}
{}.
\end{equation}
Spin charge separation is effective if the real space
separation is of order $\frac{g_4^{0}}{4 \pi \epsilon}$,
which
leads to the energy scale for spin charge decoupling:
\begin{equation}
g_4^{*} = \frac{4 \pi^{2} \epsilon}{k}
,
\end{equation}
where $k$ is the impulsion of the quasiparticle
with respect to the Fermi surface. The switching
on procedure is sucessful provided $g_4^{0} \ll g_4^{*}$.
Since the transformation (\ref{eq:45}) is independent
 on the
interactions, there is no upper limit for the rate
of switching
on $\epsilon$.

\medskip

In the same way as for the Luttinger liquid,
the sector of the
Hilbert space has to remain stable under the
action of the
evolution operator (\ref{eq:111}).
Since $U_{\epsilon}(0,-\infty)$ is diagonal in
term of charge
and spin variables, the relevant sector has a given
impulsion $k$.
This sector corresponds to a single fan of levels,
with no crossings,
except for $g_4 = 0$, leading to singular statistics.
One may compute the statistics in the whole Hilbert space,
namely to superpose the uncorrelated blocs with different impulsions.
The statistics still remain singular. The degeneracies
of some levels
are not lifted for any value of the interaction $g_4$.
These singularities
correspond to remaining degeneracies as the impulsions of
the charge and spin part are specified independently,
and are
reminiscent of the degeneracies of the free
Luttinger model.
The spectrum exhibits further singularities at non
zero level spacings, due to the linear
dependence of the levels in $g_4$: the statistics do not
become poissonian even though uncorrelated sectors are
superposed.
Note that many degeneracies, and the singularities
at non zero level
spacings are expected to disappear if $g_4$ is not
a constant as a
function of $q$. In this more generic case, the
Poisson statistics
is expected.
\section{Level spacing statistics for a model of
2 coupled chains}

We now discuss the level statistics for a model of
two coupled Luttinger liquids.
This model is solved in \cite{11} and we first remind some results.

\medskip

The two chains kinetic energy is given by:
\begin{equation}
H^{0} = v_F \sum_{k \alpha \sigma}
(k - k_F) : c_{k \alpha \sigma}^{+}
c_{k \alpha \sigma}:,
\end{equation}
where $\alpha$ labels the chain and $\sigma$ the spin.
The interactions consist of a $g_4$ term:
\begin{equation}
H^{4} = \frac{g_4}{2L}
\sum_{k \sigma \alpha}
\rho_{k \alpha \sigma}
\rho_{k \alpha -\sigma}^{+}
,
\end{equation}
and of a hopping term between the two chains:
\begin{equation}
H^{\perp} = - t_{\perp}
\sum_{k \sigma \alpha}
c_{k \alpha \sigma}^{+}
c_{k -\alpha \sigma}
{}.
\end{equation}
Only the case of two coupled one branch models is treated.
This is sufficient since no interaction couples
right and left fermions.
Fabrizio and Parola \cite{11} were able to diagonalize the hamiltonian
$H = H^{0} + H^{4} + H^{\perp}$. The excitation
spectrum of the model
exhibits four branches:
\begin{eqnarray}
\epsilon_{\rho}(q) &=& u_{\rho} q\\
\epsilon_{\sigma}(q) &=& u_{\sigma} q \\
\epsilon_{+}(q) &=&
\frac{1}{2} (u_{\rho} + u_{\sigma}) q
+
\sqrt{
(\frac{1}{2}(u_{\rho}- u_{\sigma})q)^{2}
+ 4 t_{\perp}^{2}
}\\
\epsilon_{-}(q) &=&
\frac{1}{2} (u_{\rho} + u_{\sigma}) q
-
\sqrt{
(\frac{1}{2}(u_{\rho}- u_{\sigma})q)^{2}
+ 4 t_{\perp}^{2}
}
\end{eqnarray}

The ground state is such as all the states with a
negative energy
are occupied, and all the states with a positive
energy are empty.
We computed the level statistics for a toy model
with only the
$\epsilon_{-}(q)$ branch, in a sector of given
total impulsion $q$.
We study the evolution of the statistics as the
dimensionaless
hopping constant $\tilde{t_{\perp}} =
\frac{L t_{\perp}}{\pi v_F}$
is fixed, and $\tilde{g_4} = \frac{g_4}{2 \pi v_F}$ varies.
The statistics exhibit a cross-over between two regimes as
$\tilde{g_4}$ decreases. This cross-over is
controlled by the same
lenght scale $\xi=\frac{u_{\rho}-u_{\sigma}}
{4 t_{\perp}}$ as in
\cite{11}.
If $q \xi \ll 1$, the statistics are singular,
with a sharp peak
at $s=0$. In this regime, the dispersion relation $\epsilon_{-}(q)$
may be approximated as:
\begin{equation}
\epsilon_{-}(q) = \frac{1}{2} (u_{\rho} + u_{\sigma})q
- 2 t_{\perp}
{}.
\end{equation}
The linear $q$ dependance induces high degeneracies in the excitation spectrum,
leading to a sharp peak for zero separation.

In the opposite regime ($q \xi \gg 1$), the statistics are poissonian.
The corresponding spectrum is plotted on fig.
\ref{coupled chains 1}.
In that case, the curvature of the dispersion relation
$\epsilon_{-}(q)$ is no longer negligeable, and
individual fermion
levels can no longer be considered as equidistant.

\medskip

Notice that the cross-over observed here is similar to
the case of the one dimensional, one branch Luttinger
liquid
with $q$-dependant interactions.
In both cases, the dispersion relation is linear as
the interaction parameter is set to zero (corresponding
to a highly degenerate spectrum), and becomes non linear as
interactions are switched on (leading to a random spectrum).
This transition is independent of the bosonic or
fermionic nature
of the particules. In the one dimensional Luttinger liquid,
we dealt with bosons, and the particules under
consideration in the
case of the two coupled chains are fermionic.

\medskip

What happens if we now take the four branches into account?
In the regime $q \xi \gg 1$, we observe a peak for $s=0$,
coexisting with a poissonian distribution for non
zero separations (see fig. \ref{coupled chains 2},
where the peak
is suppressed for clarity).
The peak for $s=0$ is due to the degeneracies in
the excitation
spectrum, induced by the presence of the two
linear branches.
An exemple of such degenerate configurations,
with $2$ particule
hole excitations is as follows:
the two holes have impulsions $h_1$ and $h_2$, and belong
to the $\epsilon_{-}$ branch. The particules
 with impulsions
$p_1$ and $p_2$ are on the linear
$\epsilon_{\rho}$ branch.
Consider an other excitation, deduced from
the previous one as follows:
the holes have the same impulsions
($h_1'= h_1$ and
$h_2'=h_2$). The impulsions of the
particules are such as
$p_1 + p_2 = p_1' + p_2'$.
Since all the particules belong to
the same linear branch, these
configurations
are degenerate.

Thus, the existence of the two regimes
$q \xi \gg 1$ and  $q \xi \ll 1$
in the coupled chains
is reflected in the statistical
properties of the spectrum.
To summarize, we have studied a special class
of models, since
they are integrable for any value of
the coupling constant.
In general, a non interacting fermionic
quasiparticle can be
described as a linear combination of
degenerate eigenstates,
which undergo an energy splitting as
interactions are
switched on. This is responsible for
the decay of such a quasiparticle
state, and provides a lower bound for
the switching rate $\epsilon$,
in the process of adiabatic construction
of quasiparticles.
The same degeneracy lifting has been found
to modify the energy level
spacing distribution, from a singular behaviour
for a degenerate,
non interacting system, to  a more generic
Poisson distribution already observed in
many integrable systems. We should stress
that both aspects
are non universal features of the models.
More precisely, they depends
on the complete $q$-dependance of the
interaction functions
$g_2$ and $g_4$. By contrast, universal
properties such as correlation
function exponents depend only on the
$q=0$ limit of the couplings.
We have seen that the vanishing of the
quasiparticule residue,
due to orthogonality catastrophy is also
such an universal property,
independant on the fine structure of the
spectrum and its statistics.

In this paper, we couldn't address the
question of strongly correlated
fermion systems leading to gaussian
orthogonal ensemble (G.O.E.)
statistics. However, the present work
indicates that one of the most interesting questions
is whether the difference between G.O.E.
or Poisson distribution is a universal feature of a
low-energy
fixed point or not. Our paper has been dedicated to
fine tuning phenomenas within an integrable class of
models, and the lack of universality
found here is not surprising. Intuitively, the difference
between Poisson
and G.O.E. statistics is much more robust and
might still be a way to distinguish between
several physically non equivalent fixed points.

R. M. whishes to thank
J.C. Angl\`es d'Auriac for help with
programmation, P. Degiovanni for discussions
about bosonisation and acknoledges the hospitality
of
NEC Research Institute where part of this job was
 performed.

\newpage
\renewcommand\textfraction{0}
\renewcommand\floatpagefraction{0}
\noindent {\bf Figure captions}

\begin{figure}[h]
\caption{}
\label{degeneracies}
Degeneracies of the free Luttinger model, compared to
the saddle
point approximation.
$\{\log{g(\omega)} - \frac{3}{4} \log{2} +
\frac{3}{2} \log{ 2 \pi}
+ \frac{5}{2} \log{\sqrt{\omega}}\}/\sqrt{8 \omega}$
is plotted as a function of $\sqrt{\omega}$.
This function equals $1$ for the saddle point
approximation.
In plot $(1)$,
$g(\omega)$ is the exact degeneracies. As expected,
the saddle point
approximation overevaluates the degeneracies since it
takes into
account particule-hole exitations forbidden by the
exclusion principle.
In plot $(2)$,
all the terms of the summmation
(\ref{XXX}) are taken into account. The saddle point
approximation in
(\ref{XXX}) underevaluates the degeneracies, and
becomes exact at
high energies.
\end{figure}

\begin{figure}[h]
\caption{}
\label{energy levels}
Evolution of some levels as a function of interactions.
$\varphi_q$ is a linear decreasing function of $q$,
such as $\varphi_{q=0}=2.5 a$, and
$\varphi_{q \ge 26 L/2\pi}=0$.
$a$ parametrizes the interaction strenght, and
the energy is in units of $v_F \frac{2 \pi}{L}$.
For the plot to be readable, all the levels are not
shown.
\end{figure}

\begin{figure}[h]
\caption{}
\label {convergence}
Evolution of the level spacing statistic as a function of
the cut-off $e_0$.
$\varphi_q$ is a decreasing linear function, such as
$\varphi_{q=0} = 0.25$ and $\varphi_{26\frac{2 \pi}{L}}=0$.
The statistics converge slowly to a Poissonian
distribution
("expo"). The statistics are plotted for
$e_0$ equal to $5$, $9$, $13$.
The number of levels
taken into account in the statistics is repectively:
$4196$, $97438$, $1048214$.
\end{figure}

\begin{figure}[h]
\caption{}
\label{coupled chains 1}
Level spacing statistics for the model of two coupled chains
in the regime $q \xi \gg 1$.
Only excitations of the lowest energy branch $\epsilon_{-}(q)$
are taken into account, and the analysis is restricted to
the $1p-1h$ and $2p-2h$ excitations only, for parameters
equal to: $p=200$, $t_{per} = 20$, $g=0.5$, with
the following
notations: $p$ is the total impulsion divided by $\frac{2 \pi}{L}$,
$t_{per} = \frac{L t_{\perp}}{\pi v_F}$ and $g=\frac{g_4}
{2 \pi v_F}$.
24000 states were generated. The value of the parameter
$q \xi$ is $100$.
\end{figure}

\begin{figure}[h]
\caption{}
\label{coupled chains 2}
Level spacing statistics for the model of two
coupled chains
in the regime $q \xi \gg 1$, with the four excitation branches.
Only  $1p-1h$ and $2p-2h$ excitations were taken into account.
The parameters are set to: $p=100$, $t_{per} = 100$,
$g=0.5$ and
$q \xi = 50$.
The number of computed levels is $42692$. Among them,
$8293$ separations are
equal to zero. For visibility, the level
statistics is cut off for
separations inferior as $0.05$, which supresses
the large pic at
zero separations.
\end{figure}

\end{document}